\RequirePackage{fix-cm}
\documentclass[twocolumn]{svjour3}          
\smartqed  
\usepackage{mathptmx}      
\usepackage{etex}
\usepackage{graphicx}
\usepackage{subfigure}
\usepackage{amsmath, amssymb, graphics}
\usepackage{mathtools}
\usepackage{tikz}
\usepackage{pgfplots}  
\usepackage{algorithm}
\usepackage{algorithmic}
\usepackage{pstricks}
\usepackage{booktabs}
\usetikzlibrary{shapes,arrows}
\usepackage{nomencl}
\usepackage[font=small,labelfont=bf,justification=centering]{caption}
\usepackage{placeins}
\usepackage[numbers]{natbib} 
\usepackage{stfloats}
 \journalname{Computational Mechanics}
\begin{document}

\title{Direct numerical simulation of the dynamics of sliding rough surfaces}


\author{Viet Hung DANG\and
Joel PERRET-LIAUDET, Julien SCHEIBERT, Alain LE BOT
}


\institute{Viet Hung DANG \at
Laboratoire de Tribologie et Dynamique des Systèmes, CNRS UMR 5513, \\
Ecole Centrale de Lyon, 36 Avenue Guy de Collongue, 69134 Ecully Cedex, France  \\
              Tel.: +33(0)4 72 18 62 93\\
              Fax: +33 (0)4 78 43 33 83\\
              \email{viet-hung.dang2@ec-lyon.fr}           
}

\date{Received: date / Accepted: date}

\maketitle

\begin{abstract}
The noise generated by the friction of two rough surfaces under weak contact pressure is usually called roughness noise. The underlying vibration which produces the noise stems from numerous instantaneous shocks (in the microsecond range) between surface micro-asperities. The numerical simulation of this problem using classical mechanics requires a fine discretization in both space and time. This is why the finite element method takes much CPU time. In this study, we propose an alternative numerical approach which is based on a truncated modal decomposition of the vibration, a central difference integration scheme and two algorithms for contact: The penalty algorithm and the Lagrange multiplier algorithm. Not only does it reproduce the empirical laws of vibration level versus roughness and sliding speed found experimentally but it also provides the statistical properties of local events which are not accessible by experiment. The CPU time reduction is typically a factor of 10.
\end{abstract}

\section{Introduction}
\label{SE:Introduction}
Roughness noise is the sound produced by rubbing two rough surfaces under light contact pressure~\cite{Akay2002}. This sound occurs frequently in everyday situations as hand rubbing, stridulatory sound by insects~\cite{Haskell1961} or tyre/road contact noise~\cite{Dubois2012}. In all these examples, the sound is produced by mechanical events occurring in the contact at the scale of asperities. These events may be mechanical shocks or pinning of asperities followed by a sudden release that produces a local deformation of surfaces near the contact area. The transient deformations then propagate in the solid and make it vibrate. The main features of these events are that they are rapid, numerous and unpredictable. In order to understand quantitatively the vibrational and acoustical behaviour of the system, it is necessary to understand the statistical characteristics of shocks between asperities. However the calculation of these transient dynamics is not straightforward due to the non-linearity and non-differentiability of the rough contact problem~\cite{Bussetta2012}. 

From the experimental point of view, in a series of publications~\cite{Nakai1979,Nakai1980,Nakai1981-3,Nakai1981-4,Nakai1982}, Yokoi and Nakai showed that the noise level has a strong dependence with both the surface roughness $Ra$ and sliding speed $V$. 
\begin{equation}
\Delta Lp(dB)=20\log _{10}\left[\left(\frac{{Ra}}{{R}_{{ref}}}\right)^m\left(\frac{V}{V_{{ref}}}\right)^n\right],
\label{Eq:Lv-Ra-V}
\nomenclature{$Lp$}{Sound pressure level (dB)}
\nomenclature{$Ra$}{Arithmetic average roughness$(\mu m)$}
\nomenclature{$V$}{Sliding speed $(m/s)$}
\end{equation}
where $\Delta $Lp is the increase of sound pressure level from a reference situation characterized by $R_{\text{ref}}, V_{\text{ref}}$. This law has been confirmed with various types of materials and setups by Othman and Elkholy~\cite{Othman1990} or Ben Abdelounis~\cite{Hou2010exp}. For a contact between a steel rod and a rotating disc, Yokoi and Nakai found $m$=0.8-1.2 and $n$=0.6-1.1~\cite{Nakai1982}, while with a plane-plane contact of steel surfaces, Ben Abdelounis found $m$=0.8-1.16 and $n$=0.7-0.96~\cite{Hou2010exp}. 

Boyko et al. investigated the effect of surface roughness on the noise spectrum~\cite{Stoi2007}. They found that the rougher the surface, the closer the peak in the spectra to the bending natural frequency of the system. 

Recently, Le Bot et al.~\cite{Alain2010area,LeBot2011} investigated the effect of contact area on friction sound. They found the existence of two regimes for the relationship of the noise level with contact area depending of the ratio of energy dissipated within the contact and that dissipated by the whole vibrating system. In the first regime, the noise level is constant \textit{i.e.} does not depend on the contact area, while in the second one, it is proportional to the contact area. A simple reasoning based on energy balance showed that a part of the vibrational energy is dissipated into the contact itself with a rate of dissipation $P_{diss}\propto mv^2S$ proportional to the product of the local vibrational energy $mv^2$ and the contact area $S$.

The numerical simulation of the dynamics of two frictional surfaces has been widely investigated thanks to the rapid progress of computer technology~\cite{Soom1984,Soom1986,Soom1992,Raous2001,Rigaud2003,Delogu2006,Tromborg2011,Amundsen2012}. In particular, many models have been developed to study the different types of friction noise. In Ref~\cite{Andersson2008}, Andersson and Kropp used the non-linear penalty method to compute the contacts and the Green's theory for the dynamic response of tyres in order to handle the tyre/road noise problem. In Ref~\cite{Meziane2007}, the brake noise is simulated using a finite element model and the forward increment Lagrange multiplier method. In Ref~\cite{Hou2010num}, it is shown that the relationship between the vibration level $Lv$ as a function of the surface roughness and the sliding speed can be predicted by a 2D finite element model. However, due to the micrometer size of asperities, it requires a very large number of mesh elements and this method is currently not feasible for 3D models of rough contact. 

In this paper, we propose an alternative method based on the modal decomposition to study the laws of roughness noise. The modal decomposition is an efficient method to analyze structure dynamics. The equation of motion is developed  in terms of new variables called the modal coordinates which are the solutions of a set of modal equations~\cite{Kapania1993}. The solution of the original equation is obtained via a superposition of modes. The reduction is achieved by a truncation on the mode number which has been used in the computation~\cite{Woodhouse2009}. This approach not only allows to predict the evolution of the vibration level versus $Ra$ and $V$ but also gives more quantitative and qualitative informations related to the mechanical events at the micro-asperity scale. 

The paper is organized as follows. In the next section, we describe the numerical approach which consists in the mathematical formulation, a time integration scheme and two algorithms for the contact. In Section 3, two validation tests are presented. The first one is a comparison with an analytical solution and the second one with the finite element method. In Section 4, a realistic problem is studied with an emphasis on the statistics of local events. Eventually, a conclusion is presented in Section 5.
\section{Simulation of sliding rough contact}
\label{SE:Modelisation}
\begin{figure}[!ht]
\centering
\includegraphics{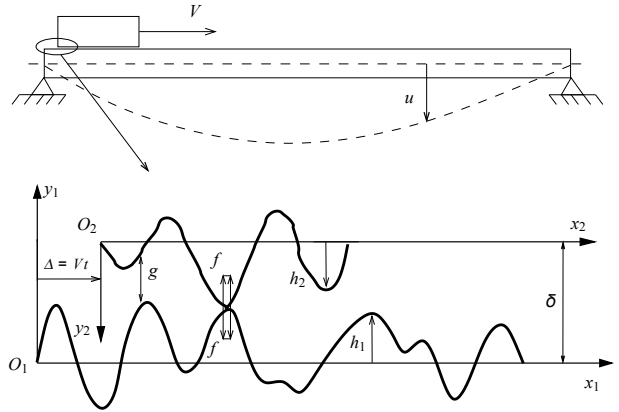}
\caption{Top: Sketch of the system under consideration. Bottom: sketch of two rough profiles. The gap between the two profiles is $\delta$. At time $t$, the horizontal offset is imposed $\Delta=Vt$.}
\label{Fi:model}
\end{figure}

The simulation is based on a 2D model which is made up of two beams in contact as shown in Fig.~\ref{Fi:model}.
The top beam moves horizontally with a constant velocity $V$ while the bottom beam is fixed at both ends. 
The beams have nominally flat rough surfaces described by their profiles. The initial vertical gap \textit{i.e.} the separation between the two reference lines of the profiles is $\delta$. During the movement asperities of the top profile can hit asperities of the bottom profile. Since a profile cannot penetrate the antagonist profile, repulsive  forces take place at contact points and lead to a global deformation $u$ of the profiles. Contacts are transient with a short duration and they result in a vibration of the whole beams. The vibrating beams then radiate sound in the surrounding air although this dissipative process is not taken into account in the present model.

We make the following assumptions:
\begin{itemize}
\item The vertical deflection $u$ of the neutral axis follows the Euler-Bernoulli theory of beams (Small flexural vibration, rotational inertia neglected).
\item The beams are infinitely rigid in the horizontal direction and therefore the horizontal position of profile nodes is imposed (No longitudinal vibration). 
\item Profiles cannot penetrate each other (Signorini's condition). 
\item The persistence of contact is ensured by a vertical gravity force.
\end{itemize}

In this model, the beams thicknesses are intended to be the actual thicknesses of the two solids under consideration. The underlying assumption is that the effective thicknesses of the rough layers are much smaller than the thicknesses of the bulk of the beams. The validity of Euler-Bernoulli's theory is well-known: The wavelength of flexural vibrations must be large compared to the thickness. This condition imposes a high frequency limit of the order of $0.1 c/H$ where $c$ is the sound speed in the material and $H$ the thickness. Beyond this limit, higher order theories such as Timoshenko's beam must be considered.

The second major assumption of this model is that longitudinal vibration has been neglected. The main reason is that in-plane motion does not contribute to the sound radiation process (only out-of-plane motion is coupled with the surrounding fluid). Furthermore, the longitudinal vibration is generally much smaller than flexural vibration. Assuming that modal energies are equal (thermal equilibrium), the ratio of longitudinal energy to flexural energy is of order of $3\sqrt{Hf/c}$. Longitudinal vibration is therefore negligible (ratio smaller than 0.1) up to a frequency of order $0.001 c/H$.

We emphasize the fact that, in our model, friction is neglected. The first possible contribution of friction is to add a horizontal component to the contact forces. Of course, this force would influence the horizontal vibration but, as we have just argued that the latter have a negligible contribution to the emitted sound, one may admit that friction does not contribute to sound radiation at first order. The second possible contribution arises from the fact that local contacts are not horizontal. Thus, the vertical projection of the friction force should in principle be taken into account to calculate the flexural vibration. However, this component is of order of $sin(\theta) .\mu.N$ where $N$ is the normal contact force, $\mu$ the friction coefficient and $\theta$ the mean slope of asperities. In most practical situations, $\mu$ is about 0.1 and $\theta$ is smaller than about 0.1, so this component is much smaller than $N$.

\subsection{Mathematical formulation}
Each profile is described in static condition by a function $h(x)$ giving the vertical position of nodes versus abscissa. But since the top profile is moving, we introduce two frames respectively attached to the bottom and top profiles. The frame $(O_1,x_1,y_1)$ is fixed, the $x_1$-axis is oriented to the right and the $y_1$-axis is oriented upward (Fig.~\ref{Fi:model}). The bottom profile $h_1(x_1)$ is given in this frame.  The frame $(O_2,x_2,y_2)$ is shifted vertically by $\delta$ and is moving horizontally rightward at speed $V$. The $x_2$-axis is oriented rightward and the $y_2$-axis is oriented downward. The top profile $h_2(x_2)$ is given in this frame. Assuming that the origins $O_1$ and $O_2$ match at $t=0$, the transformation relationships are $x_2=x_1-Vt$ and $y_2=\delta-y_1$. Due to the presence of vibration, a vertical deflection $u_i(x_i,t)$ must be superimposed to the static position $h_i(x_i)$ of nodes. At time $t$, the coordinates of a bottom node are therefore $x_1$ and $y_1=h_1(x_1)+u_1(x_1,t)$ in the fixed frame. Similarly, the coordinates of a top node are $x_2$ and $y_2=h_2(x_2)+u_2(x_2,t)$ in the moving frame. The apparent contact zone at time $t$ is $Vt \leq x_1 \leq \min(L_1,Vt+L_2)$ and  $0 \leq x_2 \leq \min(L_2,L_1-Vt)$ where $L_1$ and $L_2$ are the length of the bottom and top beams, respectively. Furthermore, the contact force per unit width (unit : N/m) is specified by a field $f_1(x_1,t)$ in the fixed frame and $f_2(x_2,t)$ in the moving frame. By the third Newton law, 
\begin{equation}
f_1(x_1,t)=f_2(x_1 - Vt,t)
\end{equation}
in the apparent contact zone.

The governing equations for the transverse motion of profiles are:
\begin{align}
D_i \Delta^2 u_i(x_i,t)+c_i \frac{\partial u_i}{\partial t}(x_i,t) + m_i \frac{\partial^2 u_i}{\partial t^2}(x_i,t) \nonumber\\
=f_i(x_i,t) \mp m_ig,
\label{Eq:governing}
\end{align}
where $i$ is the beam index (bottom beam $i=1$, top beam $i=2$ ), $D_i=E_iI_i$ the bending stiffness, $E_i$ the Young's modulus, $I_i$ the moment of inertia, $c_i$ a viscous damping coefficient and $m_i$ the mass per unit length. The gravity force per unit width is $-m_1g$ for the bottom beam but $+m_2g$ for the top beam since the moving frame is oriented downward. In Eq.~(\ref{Eq:governing}), $f$ has unit N/m and is therefore a contact force per unit width of the beams (in the direction perpendicular to the plane $(O_1 x_1 y_1)$). For simplicity, $f$ will hereafter be denoted as ''contact force''.

The boundary conditions may be either pinned-pinned without external moments ends,
\begin{equation}
u_i(0,t)=u_i(L_i,t)=\frac{\partial^2u_i}{\partial x_i^2}(0,t)=\frac{\partial^2u_i}{\partial x_i^2}(L_i,t)=0,
\label{Eq:boundary-pinned}
\end{equation}
or free-free unloaded ends,
\begin{align}
\frac{\partial^2u_i}{\partial x_i^2}(0,t)=\frac{\partial^2u_i}{\partial x_i^2}(L_i,t)=0 \nonumber \\
\frac{\partial^3u_i}{\partial x_i^3}(0,t)=\frac{\partial^3u_i}{\partial x_i^3}(L_i,t)=0.
\label{Eq:boundary-free}
\end{align}
The beams are assumed to be at rest at $t=0$ so that the initial conditions are,
\begin{equation}
u_i(x_i,0)=\frac{\partial u_i}{\partial t}(x_i,0)=0.
\end{equation}
A gap function $g$ is defined in the apparent contact zone as the vertical distance between the bottom and top profiles at any position,
\begin{align}
g(x_1,t)&=\delta - h_1(x_1) - u_1(x_1,t) \\
&- h_2(x_1-Vt) - u_2(x_1-Vt,t).
\label{Eq:gap}
\end{align}
The contact is managed through Signorini's conditions: 
\begin{equation}
g(x_1,t) \geq 0  \ ; \ f_1(x_1,t) \leq 0  \ ; \  g(x_1,t).f_1(x_1,t) = 0.
\label{Eq:Signorini}
\end{equation}
The first condition represents the impenetrability constraint while the second condition imposes the sign of contact force (repulsive forces). The third equation is the complementary condition: Either the contact force is zero (non contact) or the gap is zero (contact) but in all cases the product $f.g$ is zero. These conditions are also known as Kuhn-Tucker's conditions in the field of optimization. 
\subsection{Modal decomposition}
In this sub-section, the mathematical problem~(\ref{Eq:governing})-(\ref{Eq:Signorini}) is formulated in modal coordinates in consequence of which spatial coordinates will be removed. We introduce the uncoupled natural mode shapes $\psi_{i,k}(x_i)$:$(0,L_i)$ $\rightarrow$ $\mathbb{R}$ which are  intrinsic  properties of beam. They are  determined by solving the following problem~\cite{Karl1975},
\begin{equation}
\begin{dcases}
D_i \Delta^2 \psi_{i,k} = m_i \omega_{i,k}^2 \psi_{i,k} \\
\text{Boundary conditions~(\ref{Eq:boundary-pinned}) or~(\ref{Eq:boundary-free})}.
\nomenclature{$\omega$}{Angular eigenfrequency of beam $(rad/s)$}
\end{dcases}
\end{equation}
where $\omega_{i,k}$ is the $k$-th angular eigenfrequency of beam $i$. 

For the bottom profile (simply supported beam), the angular frequency and the mode shape functions are,
\begin{equation}
\begin{dcases}
\psi_{1,k}(x_1)=\sqrt{\frac{2}{L_1}}\sin(k+1)\frac{\pi x_1}{L_1} \\
\omega _{1,k}=\sqrt{\frac{D_1}{m_1}}\left(\frac{(k+1)\pi}{L_1}\right)^2  \quad \text{with} \quad k=0,1,2...
\end{dcases}
\end{equation}
While for the top profile (free-free beam), there is first of all two rigid-body modes to describe the vertical translation and rotation of the beam,
\begin{align}
\psi_{2,0}&=\frac{1}{\sqrt{L_2}}\\
\psi_{2,1}&=\sqrt{\frac{3}{L_2}}\frac{2}{L_2}(x-\frac{L_2}{2})
\end{align} 
and, 
\begin{equation}
\omega_{2,0}=\omega_{2,1}=0
\end{equation}
Other vibration modes are,
\begin{equation}
\begin{dcases}
\psi_{2,k}(x_2)&=\frac{1}{\sqrt{L_2}}\left[ \sin(\alpha_k x_2)+ \sinh(\alpha_k x_2)\right. + \\
&\quad+\left. \beta_k \left( \cos(\alpha_k x_2)+\cosh(\alpha_k x_2)\right) \right] \\
\omega_{2,k}&=\alpha_k^2 \sqrt{\frac{D_2}{m_2}}   \quad \text{with} \quad  k=2,3,4...
\end{dcases}
\end{equation}
where $\alpha_k$ is a modal parameter whose values are~\cite{Karl1975}: $\alpha_2=4.73/L_2, \alpha_3=7.85/L_2$ and $\alpha_k\simeq \pi[2(k-2)+3]/(2L_2)$ for $k>3$.\\
$\beta_k$ is given by:
\begin{equation}
\beta_k = \frac{\cos(\alpha_k L_2)-\cosh(\alpha_k L_2)}{\sin(\alpha_k L_2)-\sinh(\alpha_k L_2)} 
\end{equation}

Mode shapes as written above verify the orthonormality property,
\begin{equation}
\int_0^{L_i} \psi_{i,k}(x_i) \psi_{i,l}(x_i) dx_i = \delta^{kl}.
\label{Eq:orthonormality}
\end{equation}
where $\delta^{kl}$ is the Kronecker delta.

By using a modal decomposition, the transverse displacement can be written as:
\begin{equation}
u_i(x_i,t)=\sum ^\infty _{k=0}\psi_{i,k}(x_i)U_{i,k}(t),
\label{Eq:u_decomposition}
\end{equation}
and similarly for the contact force,
\begin{equation}
f_i(x_i,t)=\sum^\infty_{k=0}\psi_{i,k}(x_i) F_{i,k}(t), 
\label{Eq:f_decomposition}
\end{equation}
and the gravity force,  
\begin{equation}
m_i g=\sum^\infty_{k=0}\psi_{i,k}(x_i) G_{i,k}, 
\label{Eq:g_decomposition}
\end{equation}
where $U_{i,k}:(0,T)\rightarrow \mathbb{R}$ is the modal amplitude, $F_{i,k}:(0,T)\rightarrow \mathbb{R}$ is the modal contact force, $G_{i,k}$ is the modal gravity force and $T$ is the simulation duration. In the above three formulas, the modal component $A_{i,k}$ is given by
\begin{equation}
A_{i,k}(t)=\int^{L_i}_0 a(x_i,t) \psi_{i,k}(x_i) dx_i,
\label{Eq:modalprojection_calcul}
\end{equation}
where $A_{i,k}$ is respectively $U_{i,k},F_{i,k}$ and $G_{i,k}$ and $a$ is respectively $u$,$f$ and $m_i g$.

Substituting Eqs.~(\ref{Eq:u_decomposition})-(\ref{Eq:g_decomposition}) into Eq.~(\ref{Eq:governing}), multiplying by an arbitrary mode $\psi_{i,l}$, integrating over $x_i$ and applying the orthonormality condition~(\ref{Eq:orthonormality}) give,
\begin{equation} 
m_i \left[\ddot U_{i,k} + 2\zeta_{i,k} \omega_{i,k} \dot U_{i,k} + \omega_{i,k}^2 U_{i,k} \right] = F_{i,k}(t)\mp G_{i,k},
\label{Eq:modal-equation}
\end{equation} 
where $\zeta_{i,k}=c_i/(2 m_i \omega_{i,k})$ is the modal damping ratio. So, the new problem to solve has two unknowns the time functions $U_{i,k}(t)$ and $F_{i,k}(t)$, whose initial conditions are: 
\begin{equation}
U_{i,k}(0)=\dot U_{i,k}(0)=F_{i,k}(0)=0 \quad  
\end{equation}
with $k=0,1,2,...$ and $i=1,2$. 
At any time, the physical fields $u_i(x_i,t)$ and $f_i(x_i,t)$ are given by Eqs.~(\ref{Eq:u_decomposition}) and~(\ref{Eq:f_decomposition}) and automatically verify the boundary conditions~(\ref{Eq:boundary-pinned}) or~(\ref{Eq:boundary-free}). The contact condition~(\ref{Eq:Signorini}) to be verified requires the calculation of  $g$ through Eq.~(\ref{Eq:gap}). In principle, an exact solution of Eqs.~(\ref{Eq:governing})-(\ref{Eq:Signorini}) is obtained when using  an infinite number of natural modes. However, it would require a very small time step and a huge CPU time. Thus in practice, a truncation of the series~(\ref{Eq:u_decomposition}) and~(\ref{Eq:f_decomposition}) is made and only the first $M_i$ modes are taken into account in actual computations. For the study of roughness noise, we may restrict to modes contained in the audio range [20Hz- 20kHz]. For example for the steel beam used here, having dimension 45 $\times$ 0.2 cm, the number of modes is 30 ($f_{30}$=20900 Hz). 	
\subsection{Time integration scheme}
The second order differential equation~(\ref{Eq:modal-equation}) is solved by using the numerical scheme called leap-frog or central difference scheme. It is explicit, second order consistent, conditionally stable and simple to implement. This scheme has been successfully applied by Carpenter and al~\cite{Carpenter1991} and Meziane and al.~\cite{Meziane2007}. The time is first discretized with a constant step $\tau$, the time sequence being $t_0=0, t_1=\tau, ..., t_n=n\tau$. The modal displacement $U_{i,k}(t_n)$ defined as the exact solution of Eq.~(\ref{Eq:modal-equation}) at instant $t_n$, is approximated by the sequence $ U_{i,k,n}$ and similarly the modal velocity $\dot{U}_{i,k}(t_n)$ by $\dot U_{i,k,n}$, the modal acceleration $\ddot{U}_{i,k}(t_n)$ by $\ddot U_{i,k,n}$ and the modal force $F_{i,k}(t_n)$ by $ F_{i,k,n}$. In the central difference scheme these approximations are calculated by,
\begin{equation}
\begin{dcases}
\dot U_{i,k,n}=\frac{U_{i,k,n+1}-U_{i,k,n-1}}{2\tau},\\
\ddot U_{i,k,n}=\frac{U_{i,k,n+1}-2U_{i,k,n}+U_{i,k,n-1}}{\tau^2}.
\end{dcases}
\label{Eq:diff}
\end{equation}
Introducing Eq.~(\ref{Eq:diff}) in Eq.~(\ref{Eq:modal-equation}) gives the linear recurrence rule,
\begin{align}
U_{i,k,n+1}-
\frac{2-(\tau \omega_{i,k})^2}{1+\tau \zeta_{i,k} \omega_{i,k}} U_{i,k,n}-
\frac{\tau \zeta_{i,k} \omega_{i,k}-1}{1+\tau \zeta_{i,k} \omega_{i,k}} U_{i,k,n-1}= \nonumber\\
=\frac{\tau^2}{m_i} \left( F_{i,k,n} \mp G_{i,k}\right) .
\label{Eq:diffmodalequation}
\end{align}
Since the central difference scheme is a scheme of second order, we need to initialize the variables for the first two steps. At $t_0=0$ the displacement $U_{i,k,0}$, the velocity $\dot{U}_{i,k,0}$ are set to zero. At $t_1=\tau$, the first order explicit Euler scheme applies to calculate $U_{i,k,1}$.
\begin{equation}
U_{i,k,1}=\mp\frac{G_{i,k}}{m_i}.\frac{\tau^2}{2} 
\end{equation}
At $t_2$ and subsequently, Eq.~(\ref{Eq:diffmodalequation}) applies.

The leap-frog method is stable and convergent if $\tau < 2/\omega_{max}$ where $\omega_{max}$ is the highest value of $\omega_{i,k}$ ($k=0,1,..,M_i$) \cite{yang2005}. This time integration scheme is explicit one and the calculation of the modal amplitude at instant $t_{n+1}$ only requires the evaluation of contact forces $F_{i,k,n}$ at $t_n$. The procedure to compute $F_{i,k,n}$ will be presented in the next sub-section.
\subsection{Contact algorithm}
The horizontal position of nodes of profile $i$ is denoted $x_{i,l}=\chi. l$ with $l=1...N_i$ in frame $i$, where $N_i$ is the total number of nodes on profile. The vertical static position is noted $h_{i,l}$ while the deflection is $u_{i,l,n}$ at time $t_n$ and position $x_{i,l}$. It is obtained from $U_{i,k,n}$ by the modal composition of Eq.~(\ref{Eq:u_decomposition}),
\begin{equation}
u_{i,l,n}=\sum^{M_i}_{k=0}\psi_{i,k}(x_{i,l})U_{i,k,n},
\label{Eq:u-composition-discret}
\end{equation}
where $M_i$ is the number of modes of profile $i$ . The contact force per unit length at time $t_n$ and position $x_{i,l}$ is noted $f_{i,l,n}$. The contact condition (\ref{Eq:Signorini}) is to be applied only on the discrete set of nodes~\cite{Johnson1987} but not at other points between two nodes. 
\subsubsection{Detection of contact} 
\label{Se:detection}
For contact detection, the fast and flexible node-to-segment algorithm has been used~\cite{Zavarise2009}. The principle of the method consists of selecting a profile, called the slave profile, and checking if its nodes are in contact with the antagonist profile, called the master profile. In order to ensure the symmetry of the algorithm, the node-to-segment procedure is applied two times by exchanging the role of master and slave profiles. Such a two-pass algorithm is used to detect more efficiently the contact points. Possible artefacts of this procedure \cite{Puso2004}, related to the discontinuity of the slope of the discretized topography, are expected to be avoided through smoothing of the topography  (see Eqs. (\ref{Eq:gap_hermite}) and (\ref{Eq:hermite_equation})), as done e.g. in Ref \cite{Batailly2013}.

 Considering a slave node of abscissa $x_{i,l}$, the first step is to identify the corresponding master segment. From here, we use the symbol $i$ to denote the slave profile, $i'$ the master one with $i,i'=1,2$. Since we have assumed that the horizontal position of nodes is imposed, the master segment is found by selecting the segment which contains the vertical projection of the slave node. The master segment has index say $l'$ so that the condition 
 \begin{equation}
x_{i',l'} \leq x_{i,l} \mp V.t_n < x_{i',l'+1},
\label{Eq:mastersegment}
 \end{equation}
is fulfilled where the sign depends on the relevant change of coordinates (minus for $i=1$ and plus for $i=2$).

\begin{figure}[!ht]
\centering 
\includegraphics{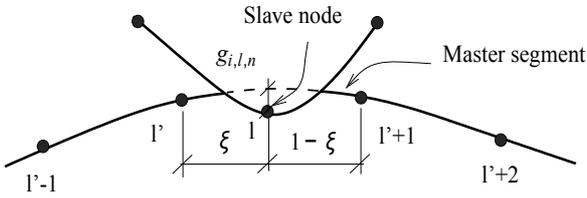}
\caption{Node-to-segment contact detection}
\label{Fi:nodesegment}
\end{figure}

 Let us introduce a local dimensionless coordinate of the vertical projection of the slave node $x_{i,l}$,
\begin{equation}
\xi = \dfrac{x_{i,l}\mp V.t_n - x_{i',l'}}{\chi}.
\label{Eq:localcoordinate}
\end{equation}  
The gap defined by Eq.~(\ref{Eq:gap}) is the vertical distance between the slave node and the interpolated master segment (Fig.~\ref{Fi:nodesegment}). Interpolation is used to ensure that the slope of the topography is everywhere continuous. We have chosen a Hermite cubic smoothing procedure to interpolate the master segments so that the gap at node $l$ on profile $i$ becomes~\cite{laursen2003},
\begin{align}
g_{i,l,n}=& \delta - (h_{i,l} + u_{i,l,n})  \nonumber \\
&-\sum_{r=0}^3 N_r(\xi)\left( h_{i',l'+r-1}+u_{i',l'+r-1,n}\right), 
\label{Eq:gap_hermite}
\end{align}
where 
\begin{equation}
\begin{dcases}
N_0(\xi)=-0.5\xi+\xi^2-0.5\xi^3,\\
N_1(\xi)=1-2.5\xi^2+1.5\xi^3, \\
N_2(\xi)=0.5\xi+2\xi^2-1.5\xi^3,\\
N_3(\xi)=-0.5\xi^2+0.5\xi^3.
\end{dcases}
\label{Eq:hermite_equation}
\end{equation}
The Hermite interpolation requires four nodes $l'-1,l',l'+1,l'+2$ and cannot be used for the two extremities ($l'=0$ and $l'=N_{i'}-1$). For these two special cases a linear interpolation is used,
\begin{align}
g_{i,l,n}= \delta - (h_{i,l} + u_{i,l,n}) - (1-\xi) \left( h_{i',l'}+u_{i',l',n}\right) \nonumber \\
 - \xi \left( h_{i',l'+1}+u_{i',l'+1,n}\right).
\end{align}
If $g_{i,l,n}<0$ then a contact is active. To avoid any miss of contact detection, the role slave/master of the two surfaces is swapped and the node-to-segment procedure is applied two times at each time step. In the next section, two methods for computing the contact forces are presented: The penalty and the Lagrange multiplier methods.
\subsubsection{Penalty method} 
In the penalty method, the contact forces are deduced from an explicit relationship with penetration, 
\begin{equation}
f_{i,l,n}=\kappa .g_{i,l,n},
\label{Eq:penaltyforce}
\end{equation}
where $\kappa$ is the penalty coefficient. By applying the equilibrium condition for the master segment (the sum of force and moment must be zero), the contact reactions at master nodes are,
\begin{equation}
f_{i',l'-1+r,n}= N_r(\xi) f_{i,l,n} \quad \text{with } r=0,1,2,3, 
\label{Eq:reaction}
\end{equation}
for $1<l'<N_{i'}-1$, and,
\begin{align}
f_{i',l',n}    = (1-\xi). f_{i,l,n}, \quad  f_{i',l'+1,n}  = \xi. f_{i,l,n} \nonumber 
\label{Eq:reaction}
\end{align}
for $l'=1$ or $N_{i'}-1$.

The contact forces are determined for all slave nodes by Eq.~(\ref{Eq:penaltyforce}) and master segments by Eq.~(\ref{Eq:reaction}). Since the role of slave and master is reversed in the second step of the contact algorithm, the contact force at any node of a profile is the sum of the penalty force and the possible reactions. The modal forces are obtained by a modal projection from Eq.~(\ref{Eq:modalprojection_calcul}). Using the trapezoidal rule yields,
\begin{equation}
F_{i,k,n}=\frac{\chi }{2}\sum _{l=1}^{N_i-1} \left(\psi _{i,k}\left(x_{i,l}\right)f_{i,l,n}+\psi _{i,k}\left(x_{i,l+1}\right)f_{i,l+1,n}\right).
\label{Eq:trapezoidal}
\end{equation}

By Eq.~(\ref{Eq:trapezoidal}), the right-hand side of Eq.~(\ref{Eq:diffmodalequation}) is well-determined at each time step. The penalty method is simple, fast and easy to implement (Algorithm 1). However the results depend on the penalty coefficient $\kappa$. A too low value of $\kappa$ causes large penetrations, while a too high value of $\kappa$ may induce unstable motion.  Following Mohammadi~\cite{discontinuum2003}, a first estimate for $\kappa$ is $0.5E < \kappa < 2.0E$ where $E$ is Young's modulus.
\begin{algorithm}[ht!]
\caption{ Penalty algorithm}
\begin{algorithmic} 
\STATE Initialize $U_{i,k,0}=0$, $U_{i,k,1}=\mp\frac{G_{i,k}}{m_i}.\frac{\tau^2}{2}$.
\STATE Loop over time steps: 
\FOR{$n=1$ to $N_T$ }
\STATE Initialize $f_{i,l,n}$
\STATE Loop for two passages algorithms:
\FOR{$i=1$ to $2$ }
\STATE compute displacement $u_{i,l,n}$ by Eq.~(\ref{Eq:u-composition-discret})
\FOR{$l=1$ to $N_i$ }
\STATE Determine master segment $l'$, local coordinate $\xi$ by Eqs.~(\ref{Eq:mastersegment}) and (\ref{Eq:localcoordinate}).
\STATE Compute gap $g_{i,l,n}$ by Eq.~(\ref{Eq:gap_hermite})
\IF {$g_{i,l,n}>0$}
\STATE $f_{i,l,n}=\kappa.g_{i,l,n}$
\STATE Compute $f_{i',l',n}$ of master surface by Eq.~(~\ref{Eq:reaction})
\ENDIF
\ENDFOR
\ENDFOR
\STATE Compute $F_{i,k,n}$ by Eq.~(\ref{Eq:trapezoidal})
\STATE Compute $U_{i,k,n+1}$, $\dot U_{i,k,n}$, go to next time step
\ENDFOR
\end{algorithmic}
\end{algorithm}

\subsubsection{Lagrange multiplier}
The second method for computing contact forces is the forward increment Lagrange multipliers~\cite{Carpenter1991,Meziane2010}. With this method, the contact forces are calculated at instant $t_n$ to satisfy exactly the non-penetration condition at time $t_{n+1}$. 

First of all, one must evaluate how a unit variation of contact force at instant $t_n$ and position $x_{i,l}$ modifies the gap elsewhere at the next instant $t_{n+1}$. Let us fix a node $l$ in slave profile $i$. A unit contact force $ \Delta f_{i,l,n}=-1$ is applied at $x_{i,l}$. The reaction forces on the master profile are
\begin{equation}
\Delta f_{i',l'-1+r,n} = - N_r(\xi) \quad \text{with } r=0,1,2,3,
\end{equation}
 where $\xi$ is the local coordinate of node $x_{i,l}$ on the corresponding master segment $l'$ of profile $i'$.
 
 The resulting variations of modal forces are obtained again by a modal projection. Applying Eq.~(\ref{Eq:trapezoidal}) with $\Delta f_{i,l,n}$ and $\Delta f_{i',l'-1+r,n}$ as above,
\begin{align}
\begin{dcases}
\Delta F_{i,k,n} = -\chi. \psi_{i,k}(x_{i,l}),\\
\Delta F_{i',k,n} =- \chi.  \sum^3_{r=0} N_r(\xi).\psi_{i',k}(x_{i',l'-1+r}). 
\end{dcases}
\label{Eq:variation-F}
\end{align}
The variation of modal amplitudes of the slave profile at time $t_{n+1}$ which results from this variation of modal forces is obtained by Eq.~(\ref{Eq:diffmodalequation}).
\begin{equation}
\Delta U_{i,k,n+1}= \frac{\tau^2}{m_i} \Delta F_{i,k,n},
\label{Eq:variation_U}
\end{equation}
and similarly $\Delta U_{i',k,n+1}$ for the master profile.

 By applying Eq.~(\ref{Eq:u-composition-discret}), the variation of deflection at any point $x_{i,m}$ on the slave profile is
\begin{equation}
\Delta u_{i,m,n+1}=\sum _{k=0}^{M_i} \psi _{i,k}\left(x_{i,m}\right)\Delta U_{i,k,n+1},
\label{Eq:variation_petit_u}
\end{equation}
and similarly for the master profile.

Eq.~(\ref{Eq:gap_hermite}) gives the variation in gap at any point $x_{i,m}$ on profile $i$ caused by a unit contact force applied at $x_{i,l}$ ,
\begin{equation}
\Delta g_{i,m,n+1}=- \Delta u_{i,m,n+1} - \sum^3_{r=0}N_r(\xi) \Delta u_{i',m'-1+r,n+1} 
\label{Eq:Gij}
\end{equation}
where $m',m'+1$ is the master segment containing the vertical projection of slave node $m$.

Let us introduce the influence matrix $[\Delta]$ of size $N_i \times N_i$ whose component $\Delta_{m,l}$ is the variation of gap  at node $x_{i,m}$ on profile $i$ due to a unit contact force $ \Delta f_{i,l,n}=-1 $ applied at $x_{i,l}$ on slave profile $i$. We have
\begin{equation}
\Delta_{m,l} = \Delta g_{i,m,n+1}.
\end{equation}
Then the computational procedure is the following. The contact force $f_{i,l,n}$ is initialized to $f^0_{i,l,n}$ which may be either 0 or $f_{i,l,n-1}$. Then the gap $g^0_{i,l,n+1}$ is predicted by Eqs.~(\ref{Eq:diffmodalequation}),(\ref{Eq:u-composition-discret}) and (\ref{Eq:gap_hermite}).  Detection of penetration at instant $t_{n+1}$ is then realized. We denote $N_c$ the number of predicted penetration nodes. In practice, this number is much smaller than the number of nodes $N_c<<N_i$~\cite{Greenwood1966}. Let us denote $\{q\}$ the vector containing the indices of penetrating nodes in increasing order $q_{\alpha} < q_{\alpha+1}$ and $\{e\}$ the $N_c$ penetration values. If $q_\alpha=l$ then $e_{\alpha}=g_{i,l,n+1}^0$. The effective influence matrix $[\Delta ']$ of dimension $N_c\times N_c$ is extracted from the influence matrix $[\Delta]$,
\begin{equation}
\Delta '_{\alpha,\beta}=\Delta_{q_\alpha,q_\beta}.
\end{equation}

Now, we introduce the vector of Lagrange multipliers $\{\lambda\}$ which is the variation of contact force to apply at penetrating node $\lambda_\alpha=\Delta f_{i,q_{\alpha},n}$. In order to cancel the penetration, we solve the following linear system equation,
\begin{equation}
-\{e\}=[\Delta '].\{\lambda\}.
\label{Eq:systemequation}
\end{equation}
 The obtained solution $\lambda$ is used to correct the calculation of contact force:
 \begin{equation}
f_{i,q_{\alpha},n}=f^0_{i,q_{\alpha},n}+\lambda_\alpha.
\end{equation}

However, it is not ensured that all these contact forces are non-positive. Furthermore, we have imposed the gap to be zero only in the formerly detected contact zone. But it may happen that new penetrating nodes appear with $f_{i,q_{\alpha},n}$. So, all the above steps are implemented in an iterative algorithm (see Algorithm 2). The contact force $f^0_{i,l,n}$ is updated with the last value $f_{i,l,n}$. At each step, the iterative process stops when the gap is non-negative everywhere and the contact forces are all non-positive. In practice, the number of penetrated nodes is relatively small compared with the total number of nodes and the distance between contact asperities is quite large. In these conditions, only one iteration of the algorithm is sufficient to provide  accurate results.

 The main advantage of the Lagrange multipliers method is that the contact condition is satisfied exactly, without requiring the use of any empirical parameter contrary to the penalty method. However, it introduces a set of unknown variables and extra equations associated with the Lagrange multipliers. As a result, the computational procedure is more complex and takes more CPU time.
\begin{algorithm}[ht!]
\caption{ Lagrange multiplier algorithm}
\begin{algorithmic} 
\STATE Initialize $U_{i,k,0}=0$, $\dot U_{i,k,0}=0$, $U_{i,k,1}=\mp\frac{G_{i,k}}{m_i}.\frac{\tau^2}{2} $.
\STATE Loop over time steps: 
\FOR{$n=1$ to $N_T$ }
\STATE Suppose $f^0_{i,l,n}=0$, predict $U^0_{i,k,n+1}$, $u^0_{i,l,n}$ by Eqs.~(\ref{Eq:diffmodalequation}) and~(\ref{Eq:u-composition-discret}).
\STATE Loop for two passages algorithms:
\FOR{$i=1$ to $2$ }
\STATE Compute influence matrix $[G]$ by Eqs.~(\ref{Eq:variation-F})-(\ref{Eq:Gij})
\STATE Gauss-Seidel iterations :
\FOR {j = 0, 1, 2, . . . ,till convergence }
\FOR{$l=1$ to $N_i$ }
\STATE Determine master segment $l'$, local coordinate $\xi$ by Eqs.~(\ref{Eq:mastersegment}), (\ref{Eq:localcoordinate}).
\STATE Compute gap $g^j_{i,l,n+1}$ by Eq.~(\ref{Eq:gap_hermite})
\STATE Initialize number of penetration node $N_c=0$;
\IF {$g^j	_{i,l,n+1}>0$}
\STATE $N_c=N_c+1, \quad e_{N_c}=g^j_{i,l,n+1}, \quad q_{N_c}=l, \quad q'_{N_c}=l'$ \ENDIF
\ENDFOR
\STATE Compute effective influence matrix $E_{\alpha,\beta}=G_{q_{\alpha},q_{\beta}}$ 
\STATE Solve linear equations: $[E].\{ \lambda\}=\{e\}$
\STATE Correct contact force
\FOR{$\alpha=1$ to $N_c$ }
\IF {$f^j_{q_{\alpha}}+\lambda_{\alpha} < 0$} 
\STATE $f^{j+1}_{q_{\alpha}}=f^j_{q_{\alpha}}+\lambda_{\alpha}$
\STATE Compute reaction forces of master surface by Eq.~(\ref{Eq:reaction})
\ENDIF
\ENDFOR
\STATE Correct gap $ \{g^{j+1}_{i,n+1}\}=  [G].\{f^{j+1}_{i,n}\}$
\IF {non-penetration condition of $\{g^{j+1}_{i,n+1}\}$ is tolerable}
\STATE Leave Gauss-Seidel iteration
\ELSE
\STATE Go to next Gauss-Seidel iteration.
\ENDIF
\ENDFOR
\ENDFOR
\STATE Corrector $U_{i,k,n+1}$, compute $\dot U_{i,k,n}$, go to next time step
\ENDFOR
\end{algorithmic}
\end{algorithm}

\section{Validation tests}
\begin{figure}[!ht]
  \centering 
\includegraphics{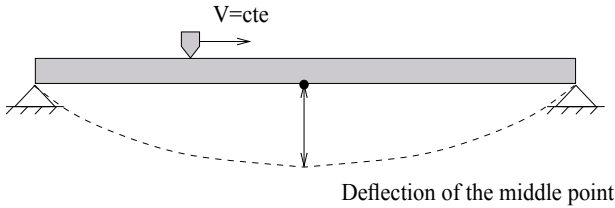}  
\caption{Moving mass problem}
\label{Fi:model-Olsson}
\end{figure}

The algorithms presented above have been implemented in the program RA2D written in C language. Two numerical tests are presented in this section in order to validate the accuracy of the proposed approach. In the first test, we compare RA2D with an analytical solution of  the problem of a mass moving on a flexible horizontal beam. This problem is common in civil engineering applications when structures such as bridges, rails or roadways are subjected to vehicles load. The model consists of a moving mass with a constant horizontal velocity on an Euler-Bernoulli beam (see Fig.~\ref{Fi:model-Olsson}). The moving mass is a rigid solid of mass $M$=0.36 kg. The beam has length $L$=11 m, Young's modulus $E=$1.7e8 N/m$^2$, density $\rho$=3100 kg/m$^3$, cross sectional area $A=$0.005 m$^2$ and inertia $I=$2.6e-5 m$^4$. The number of modes of the beams taken into account is 20. The moving mass is modelled with the sole two rigid modes (vertical translation and rotation, the horizontal position being imposed), the action of gravity and the contact force. The input and simulation parameters are given in Tables~\ref{Ta:input_olsson} and \ref{Ta:simulation_olsson}, respectively.
\begin{table}[!ht]%
\centering
\caption{Input parameters of the moving mass problem}
\label{Ta:input_olsson}
\begin{tabular}{ccccccccccccc}
\toprule
Mass  & $L$ & $E$ & $\rho$  & $A$ & $I$ \\
$(kg)$& $(m)$ & $(N/m^2)$& $(kg/m^3)$ & $(m^2)$ & $(m^4)$ \\
\hline
0.36   & 11.6 & 1.7e8 & 3100 & 0.005 & 2.6e-5 \\ 
\bottomrule
\end{tabular}
\end{table}
\begin{table}[!ht]%
\centering
\caption{Simulation parameters of the moving mass problem}
\label{Ta:simulation_olsson}
\begin{tabular}{ccccccc}
\toprule
 Number of & Space step & Time step & Duration   & $T_1$  \\ 
 beam modes & $(m)$&  $(s)$&   $(s)$& $(s)$\\
\hline
 20 & 1e-2   &  5e-6 & 10   & 5.11 \\ 
\bottomrule
\end{tabular}
\end{table}

This problem was analytically solved by Olsson~\cite{Olsson1991} who found a closed-form solution to the governing equation as follows:
\begin{align}
u(x,t)&=\frac{2PL^3}{\pi ^4 EI}\left[\frac{1}{2\alpha ^4}\sin \left(\frac{{\alpha \pi x}}{L}\right) \right. \nonumber\\ 
&\left. \left(\sin \left(\frac{{\alpha \pi t}}{\tau }\right)-\frac{{\alpha \pi t}}{\tau }\cos \left(\frac{{\alpha \pi t}}{\tau }\right)\right)\right]+
\nonumber\\
&+\frac{2{PL}^3}{\pi ^4{EI}}\sum _{n=1,n\neq \alpha }^{\infty } \left[\frac{1}{n^2\left(n^2-\alpha ^2\right)}\sin \left(\frac{{n\pi x}}{L}\right) \right. \nonumber\\ 
&\left. \left(\sin \left(\frac{{n\pi t}}{\tau }\right)-\frac{\alpha }{n}\sin \left(\frac{n^2{\pi t}}{\alpha \tau }\right)\right)\right].
\end{align}
where $P$ is the weight of the moving mass, $L$ the beam length, $EI$ the bending stiffness, $\tau=L/V$ the traversing time of the moving mass, $\alpha$ a dimensionless parameter characterizing the velocity of the moving mass defined by $\alpha = T_1/(2\tau)$, where $T_1=2L^2/\pi\sqrt{\rho A/(EI)}$ is the period of the first eigenmode of the beam. 
\begin{table}[!ht]%
\centering
\caption{Moving speed of the mass}
\label{Ta:movingspeed}
\begin{tabular}{cccccc}
\toprule
$V$ $(m/s)$& 0.57& 1.15& 2.3& 4.6 \\
\hline
$\alpha$ & 0.125& 0.25& 0.5& 1.0 \\
\bottomrule
\end{tabular}
\end{table}

\begin{figure}[!ht]
   \centering 
 \includegraphics{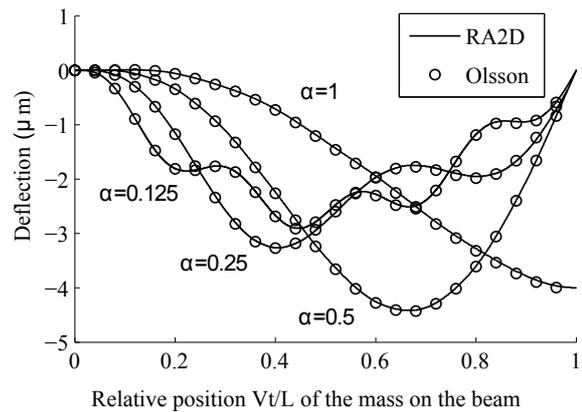}  
 \caption{Comparison of the deflection at the middle point of the beam obtained either with RA2D or with Olsson's formula, for four values of the sliding speed ($\alpha$=.125, 0.25, 0.5, 1.0). }
 \label{Fi:ComparaisonMassemobile}
 \end{figure}

 In Fig.~\ref{Fi:ComparaisonMassemobile} is shown the time evolution of the beam deflection at its middle point for four values of $\alpha$ (see Table~\ref{Ta:movingspeed}) obtained by the analytical solution~\cite{Olsson1991} and by the numerical simulation with RA2D/Lagrange. A perfect agreement between the results is observed. We observe at most 4 periods of vibration which validates a posteriori the choice of 20 modes to get satisfactory results. The study concerning the influence of number of modes on results will be done in the next section.
  
\begin{figure}[!ht]
   \centering 
    \includegraphics{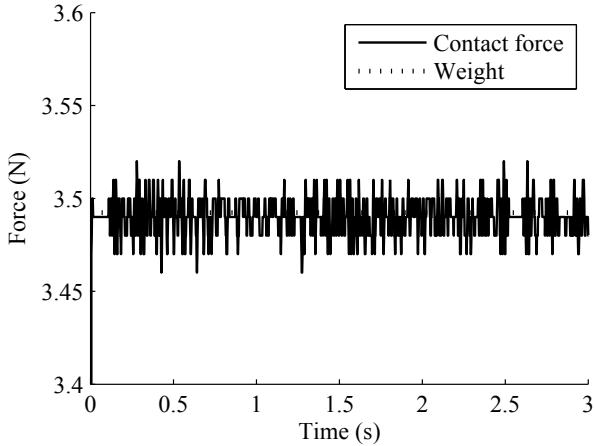}  
 \caption{Contact force obtained by RA2D with $V=0.57$ m/s ($\alpha=0.125$).}
 \label{Fi:forceComparaisonMassemobile}
 \end{figure}

 There is contact between mass and beam during the whole simulation and the numerical contact force is at all times very close to the weight of the moving mass as shown in Fig.~\ref{Fi:forceComparaisonMassemobile}. The fluctuations of the contact force can be reduced by using a finer space and time discretization steps and a higher number of modes. This first test validates the dynamical part of our code.
 
\begin{figure}[!ht]
\centering 
\includegraphics{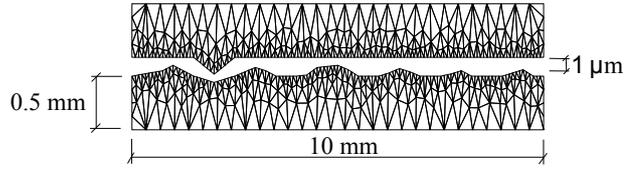}  
\caption{Mesh in ABAQUS of the contact problem between a single asperity surface (top) and a six asperities surface (bottom). Note that two different scales are used for horizontal and vertical axis.}
\label{Fi:maillagesixchocs}
\end{figure}

The second test concerns the contact part of the code. It is a toy model formed by two simple rough surfaces rubbed together. It is used to compare the program RA2D with the finite element software ABAQUS Explicit. The two beams have same dimensions, material properties and both have pinned ends. The top profile consists of only one asperity whereas the bottom profile consists of six asperities as illustrated in Fig.~\ref{Fi:maillagesixchocs}. The height of the asperities is of the order of 1 $\mu$m. A mesh with 2874 nodes and 850 CPE6 (6-node triangular plane strain) elements is used in the ABAQUS model. The space step $\chi$ is 80 $\mu m$, the time step $\tau=0.02$ $\mu s$. The input and simulation parameters are given in Tables~\ref{Ta:input_ABQ} and~\ref{Ta:simulation_input_ABQ}, respectively. 
\begin{table}[!ht]%
\centering
\caption{Input parameters of the simple asperity problem}
\label{Ta:input_ABQ}
\begin{tabular}{ccccc}
\toprule
$L$&  $H$ & $E$& $\rho$  & Damping \\
$(m)$ & $(m)$ & $(N/m^2)$ & $(kg/m^3)$ & \\
\hline
0.01& 5e-4& 5e10 & 2000  & 0\% \\
\bottomrule
\end{tabular}
\end{table}

\begin{figure*}[!ht]
\centering 
\includegraphics{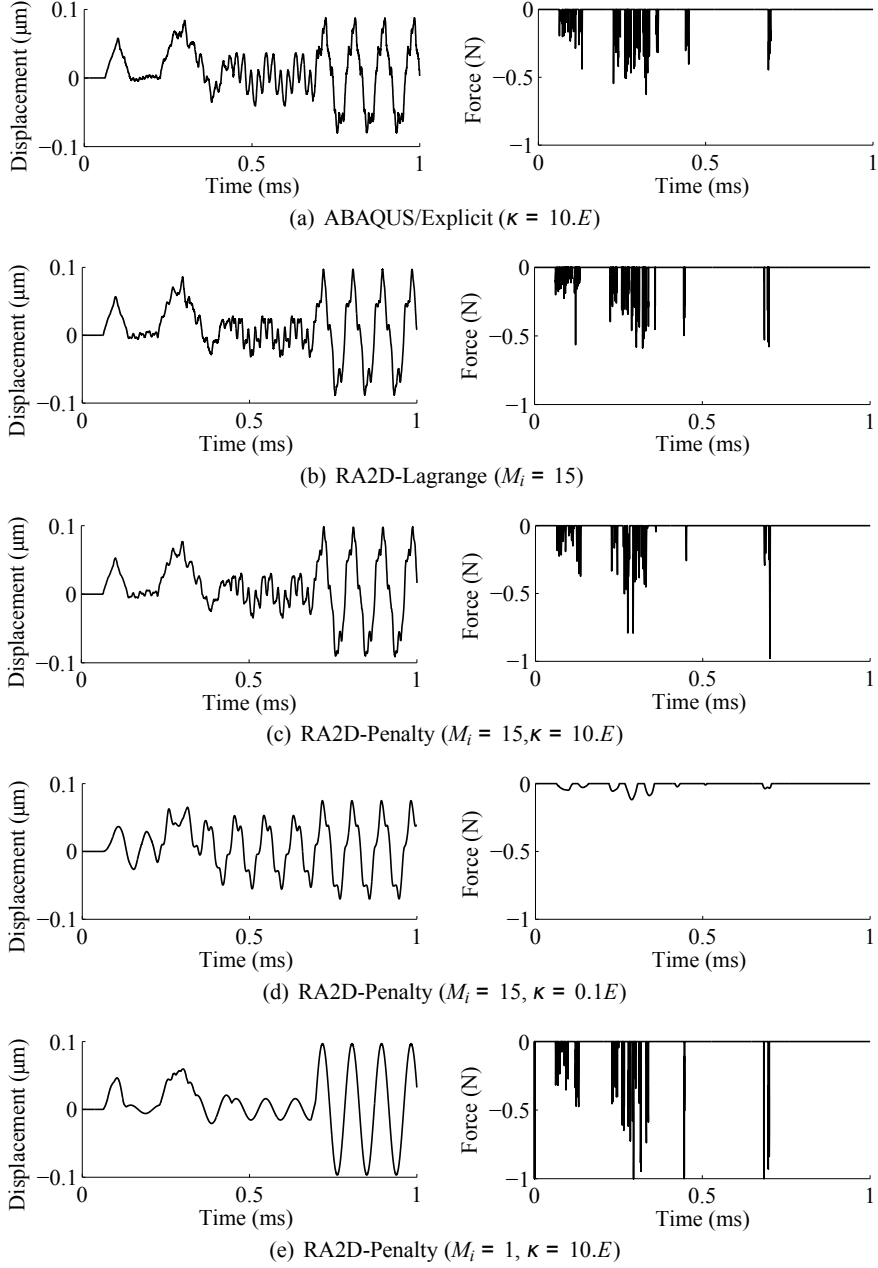}  
\caption{Comparison of displacement (left) and contact force (right) at the summit of the unique asperity of the top surface, for ABAQUS (a) or Ra2D (b to e) with different simulation parameters}
  \label{Fi:resultComparaisonABQ}  
\end{figure*}

\begin{figure}[!ht]
     \centering 
\includegraphics{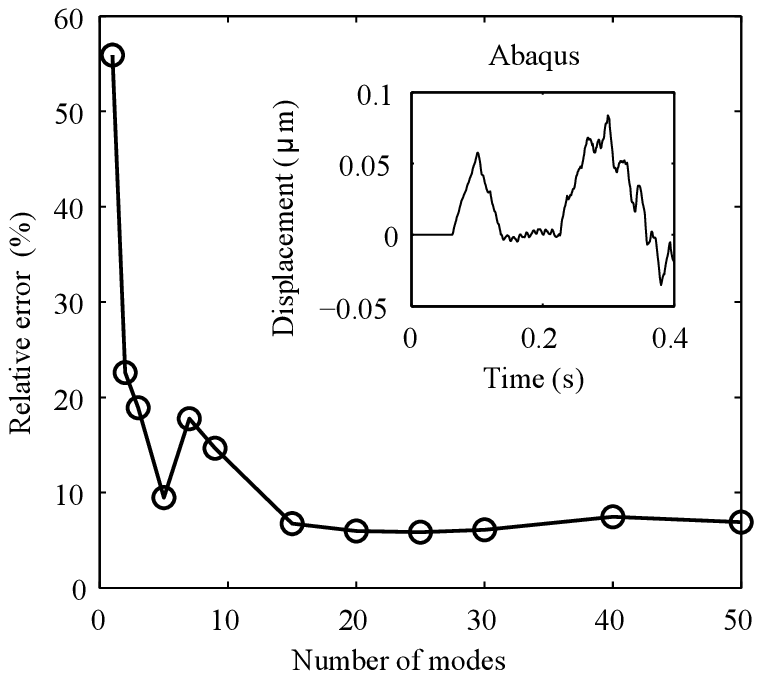}  
    \caption{Evolution of the relative error of the displacement at the summit of the top asperity between ABAQUS and RA2D as a function of the number of modes. The displacement from 0 to 0.4 s is used for the calculation of the relative error (see the inset).}
      \label{Fi:u-conververgence}  
    \end{figure}

In Fig.~\ref{Fi:resultComparaisonABQ} is shown the comparison of displacement and contact force at the summit of the unique asperity of the top profile. The results obtained by ABAQUS Explicit with the penalty method with a penalty coefficient $\kappa=10E$ are shown in Fig.~\ref{Fi:resultComparaisonABQ}a. They are taken as the reference. The multiplier Lagrange method with 15 modes shows a good agreement in the displacement evolution (Fig.~\ref{Fi:resultComparaisonABQ}b). The results are equally good using the penalty method $\kappa = 10E$ and 15 modes as shown in Fig.~\ref{Fi:resultComparaisonABQ}c.  Fig.~\ref{Fi:resultComparaisonABQ}d shows the poor results obtained by RA2D/penalty with a too low penalty coefficient ($\kappa=0.1E$). The contact force is too low, the shock duration is too large. Figure~\ref{Fi:resultComparaisonABQ}e highlights the insufficiency of a calculation with a low number of modes ($M_i$=1). The force is overestimated and all high frequency details of vibration are filtered out. In order to quantify the quality of the results, the relative error of the displacement at the summit of the asperity on the top profile between both programs RA2D and ABAQUS is calculated. The relative error is given by the RMS-value (time average) of the difference of displacements divided by the RMS-value of the displacement obtained by ABAQUS. The displacement from 0 to 0.4 s is used for the calculation of the relative error. In Fig.~\ref{Fi:u-conververgence}, the evolution of the relative error as a function of the number of modes is plotted. It is quite large for the first 10 modes while increasing the number of modes reduces it. When more than 15 modes are used, the error becomes stable and around 6\%. This residual error comes from a slight time shift between the ABAQUS and RA2D results which can be explained by (i) the fact that the contact problem is highly non-linear and as such, is very sensitive to initial conditions, numerical errors (round-off error, local truncation error) and (ii) the insufficiency of Euler-Bernoulli theory.  The ratio $L/H$ is 20 which appears to be sufficiently high for Euler-Bernoulli's theory to be applied. However, the wavelength for the highest natural frequency used, $f_{15}$, is estimated to be 1 mm. The ratio wavelength to thickness is only 2 which may explain discrepancies between elasto-dynamic and Euler-Bernoulli predictions.
\begin{table*}[!ht]%
\centering
\caption{Simulation parameters of the simple asperity problem}
\label{Ta:simulation_input_ABQ}
\begin{tabular}{cccccccc}
\toprule
 $V$ & Number of & $\chi$ & $\tau$ & Duration& $\delta $ & $f_1$  & $f_{15}$\\
 $(m/s)$ & modes &  $(m)$&  $(s)$&  $(s)$&$(m)$& $(Hz)$ & $(Hz)$ \\
\hline
 1 & 15 & 8e-5   &  1e-8 & 1e-3 & 1.45e-6 & 1.13e4 & 2.55e6\\ 
\bottomrule
\end{tabular}
\end{table*}
    
   In terms of CPU time, it takes 280 s with ABAQUS but only 30 s with RA2D/Penalty on the same computer. RA2D/Penalty is thus around ten times faster. 

\section{Realistic problem}	
\label{SE:Results}
\begin{figure}[!ht]
\centering
\includegraphics{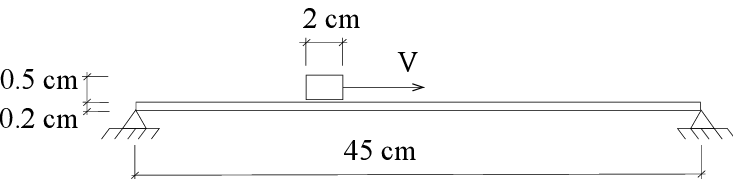} 
\caption{Realistic model}
\label{Fi:realistic_model}
\end{figure}

A realistic sliding contact problem between two solids with rough surfaces is outlined in this section. The system is made of two solids, a parallelepipedic solid moving on a simply supported Euler beam \ref{Fi:realistic_model}. The slider (top solid) is moving in the $x$-direction at constant speed $V$ in the range 0.02 $\leq V \leq$ 0.7 m/s. The resonator (bottom solid) has length 450 mm and thickness 2 mm while the slider has length 20 mm and thickness 5 mm. Both solids are made of steel with Young's modulus $E=210$ GPa, Poisson's ratio $\nu=0.3$, mass density $\rho = 7800$ kg/m$^3$ and modal damping ratio $\zeta=0.02$ for all modes.
\begin{figure}[!ht]
\centering
\includegraphics{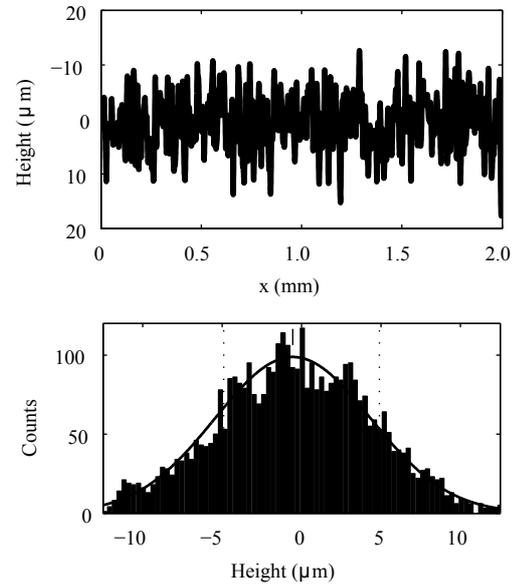} 
\caption{Numerical rough surface Ra5 (only 2 mm are shown) and its height distribution. The curve is a Gaussian fit, vertical dashed lines indicate $\pm$ one standard deviation.}          
\label{Fi:Numerical_surface}
\end{figure}

 The numerical simulations are performed with the following parameters: time step $\tau=0.1$ $\mu s$, duration of simulation $T=1$ s. All the input parameters are given in Tables~\ref{Ta:input-slider-resonator} and~\ref{Ta:simulation_input_sldier-resonator}. The rough surfaces are numerically generated by using the Garcia and Stoll's method~\cite{Garcia1984} and Bergstrom's program~\cite{Bergstrom2007} leading to surfaces having a roughness $Ra$ from 3 to 30 $\mu m$. The space step is $\chi=5$ $\mu m$. These surfaces are characterized by their standard statistical parameter $Ra,Rq,Rsk$ and $Rku$ for respectively arithmetic roughness, quadratic roughness, Skewness and Kurtosis (Table~\ref{Ta:Rough-surface}). In addition, the auto-correlation function (ACF) describes the manner in which the height varies along the surface. The correlation length $l_c$ is defined by the value where the ACF reduces to 0.37 times its value at origin. Fig.~\ref{Fi:Numerical_surface} shows an example of slider surface with roughness $Ra$=5 $\mu m$. The results presented in this section are obtained using the penalty algorithm with a penalty coefficient $\kappa$=2.1e12 $Pa$. In general, to determine the appropriate penalty coefficient, we realize a short duration simulation ($T=0.001$ s) by two algorithms: Lagrange multipliers algorithm and penalty algorithm. The results obtained by the two algorithms are then compared. If the difference of the comparison is too large ($>$10\%), the penalty coefficient is modified. 
\begin{table*}[!ht]%
\centering
\caption{Input parameters of the realistic problem}
\label{Ta:input-slider-resonator}
\begin{tabular}{cccccccccc}
\toprule
\multicolumn{3}{c}{Material} &&\multicolumn{2}{c}{Resonator}&&\multicolumn{3}{c}{Slider}\\
\cmidrule{1-3}\cmidrule{5-6}\cmidrule{8-10}$E$ & $\rho$ & $\zeta$& & $H$ & $L$ && $H$ & $L$ & Speed \\
  $(Pa)$ &$(kg/m^3)$&&&$(m)$&$(m)$&&$(m)$&$(m)$&$(m/s)$\\
\midrule
$210E9$&7800&2\% &&0.002&0.45&&0.005&0.02&$0.02 - 0.7$\\
\bottomrule
\end{tabular}
\end{table*}
\begin{table*}[!ht]%
\centering
\caption{Simulation parameters of the realistic problem}
\label{Ta:simulation_input_sldier-resonator}
\begin{tabular}{ccccccc}
\toprule
$\chi$ & $\tau$ & Duration & Number of  & $T_1$& $T_{40}$&CPU time\\
$(m)$  &  $(s)$&  $(s)$ & mode &   $(s)$ & $(s)$&($s$)\\ 
\hline
5e-6   &  1e-7 & 5 & 40  & 0.034& 2.12e-5 & 9920 \\ 
\bottomrule
\end{tabular}
\end{table*}
\begin{table}[!ht]%
\centering
\caption{Numerical rough surface}
\label{Ta:Rough-surface}
\begin{tabular}{ccccccc}
\toprule
Profile  & $Ra$ & $Rq$ & $Rsk$ & $Rku$ & $l_c$  \\
		 & $(\mu m)$ & $(\mu m)$ & $(\mu m)$ & $(\mu m)$ & $(\mu m)$  
\\\midrule
Ra3  & 2.89& 3.57 & -0.04 & 3.07 & 400 
\\\midrule
Ra5  & 4.86& 6.02& -0.12 & 3.33 & 450
\\\midrule
Ra8 & 7.72 & 9.65 & 0.05 & 3.10 & 450
\\\midrule
Ra10& 9.54 & 11.88 & 0.08 & 3.15 & 500
\\\midrule	
Ra20& 20.39 & 25.80 & -0.01 & 3.33 & 500
\\\midrule
Ra30& 30.91 & 38.61 & -0.11 & 3.06 & 500
\\\bottomrule
\end{tabular}
\end{table}

\begin{figure}[!ht]
\centering
\includegraphics{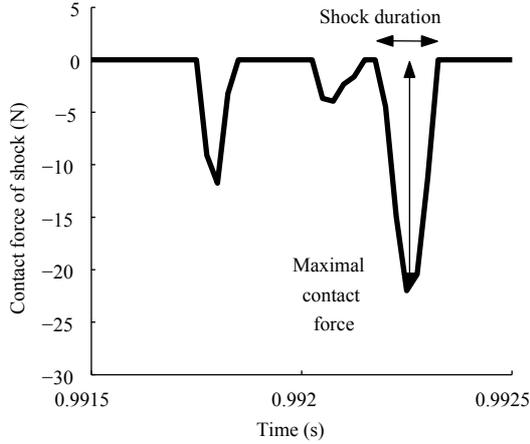}
\caption{Evolution of contact force versus time at point $x$=0.012 m on the top surface ($Ra$=5 $\mu$m, $V$=0.1 m/s). From this evolution, the maximal absolute value of contact force and shock duration $\Delta t$ are determined}
\label{Fi:shock-definition}
\end{figure}

By recording the dynamical response during the process with a specified sampling frequency, one can access the statistics  of the contact events at all nodes of the surfaces. The shock between one node with the antagonist surface is determined mathematically from the time evolution of the contact force of this node as shown in Fig.~\ref{Fi:shock-definition}. When the contact force is non-zero, the shock occurs.
\begin{figure*}[b]
 \centering 
\includegraphics{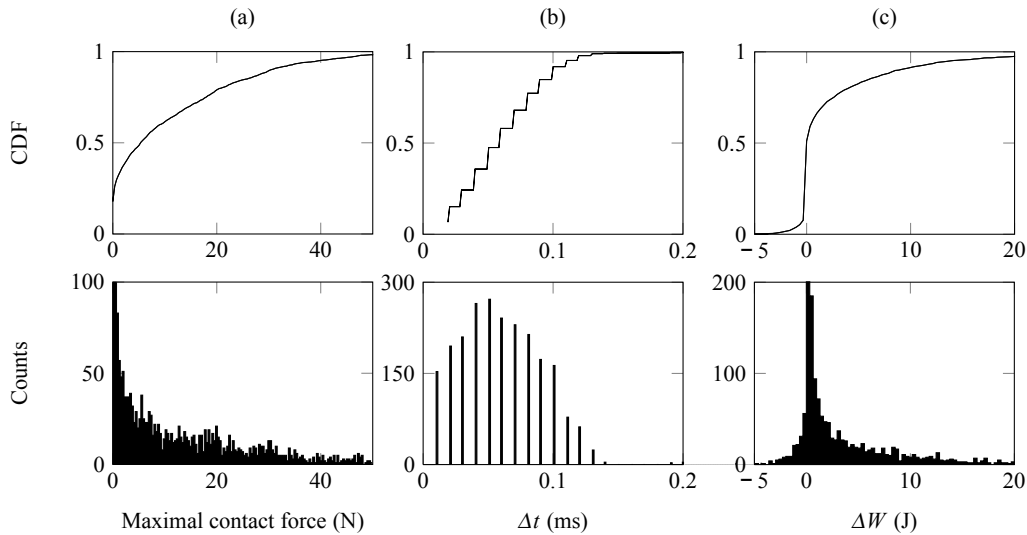}
\caption{CDF (top) and histogram (bottom) of shock properties ($Ra$= 5 $\mu$m, $V$=70 cm/s). (a) Maximal contact force of shock, (b) Shock duration, (c) Transferred energy through the shock.}
  \label{Fi:choc-implicite}  
\end{figure*}

 A shock may be caracterized by three properties: the shock duration $\Delta t$, the maximal absolute value of contact force and the transferred energy $\Delta W$. This energy is given by the formula
\begin{equation}
\Delta W = \sum_{n\in shock} \chi f_{i,l,n} \dot{u}_{i,l,n}.\tau
\nomenclature{$\Delta W$}{Energy being transferred $(J)$}.
\end{equation}
where $f_{i,l,n}$ and $\dot{u}_{i,l,n}$ are respectively the contact force and vibrational velocity at node $x_{i
,l}$ of profile $i$ where the shock occurs. The sum runs over all time steps of the shock.

In Fig.~\ref{Fi:choc-implicite} are presented typical histograms and the cumulative distribution functions (CDF) of $\Delta t$, the maximal force and $\Delta W$, on the example of $V$=70 $cm/s$ and $Ra$=5 $\mu m$. Taking the weight of the slider as the reference value ($M$=0.78 N), almost $30\%$ of the maximal absolute value of contact force is smaller than $M$, $57\%$ is smaller than 10 $M$ and $100\%$ is smaller than 100 $M$. More than $90\%$ of the shock durations are shorter than 1e-4 s (Fig.~\ref{Fi:choc-implicite}b), which is very short compared to the period of the first eigenmode of the resonator ($T_1$=0.034 s). The histogram and the CDF of $\Delta W$ are presented in Fig.~\ref{Fi:choc-implicite}c. It is observed that $\Delta W$ can be either positive or negative. This means that the contact plays a double role. When $\Delta W>0$, shocks are injecting vibrational energy into the resonator, and thus act as noise sources. They transform the kinetic energy of the slider into vibrational energy. Conversely when $\Delta W<0$, energy is transferred from the resonator to the slider. For the resonator, it is a dissipation. However, the sum of energies being transferred to the resonator through all shocks remains positive. 

\begin{figure}[!ht]
\centering
\includegraphics{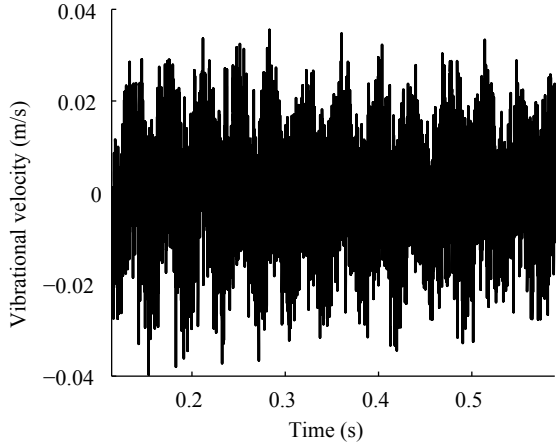}
\caption{Time evolution of vibrational velocity and its power spectral density (PSD) at point x=0.165m on the resonator for Ra=5$\mu m$ and V=0.1$m/s$. The dotted lines present the natural frequencies of the resonator ($f_4, f_7, f_{12}, f_{15}$ and $f_{18}$).} 
\label{Fi:vibration-velocity}
\end{figure}

In order to analyze the dynamical response of the resonator to the asperity shocks, we focus on the vibration velocity of discretized nodes of surfaces. An example of the time evolution of vibrational velocity and its  power spectral density (PSD) at node $x=0.165$ m on the resonator is illustrated in Fig.~\ref{Fi:vibration-velocity} for the case $V=0.1$ m/s and $Ra$=5 $\mu$m. The dotted lines present the eigenfrequencies of the resonator. The peaks in the PSD are found close to the natural frequencies of the resonator which means that the shocks essentially act as a source of excitation of the resonator, over a wide spectrum of frequencies. The fact that the frequencies are unchanged also means that the coupling between the two profiles is weak. 

The vibration level $Lv$ of the resonator is calculated using the following formula:
\begin{equation}
Lv = 20\log_{10}\left( \frac{v_{rms}}{v_{ref}}\right).
\end{equation}
where $v_{ref}$ is a reference value $v_{ref}$=1e-9 m/s \cite{norton2003} and $v_{rms}$ is the root mean square value of vibration velocity (average over both space and time) by 
\begin{equation}
v_{rms}^2=\frac{1}{T}\frac{1}{L_i}\int^T_0 \int^{L_i}_0 \dot{u}_i^2(x_i,t).dx_i.dt
\end{equation} 
\begin{figure}[!ht]
\includegraphics{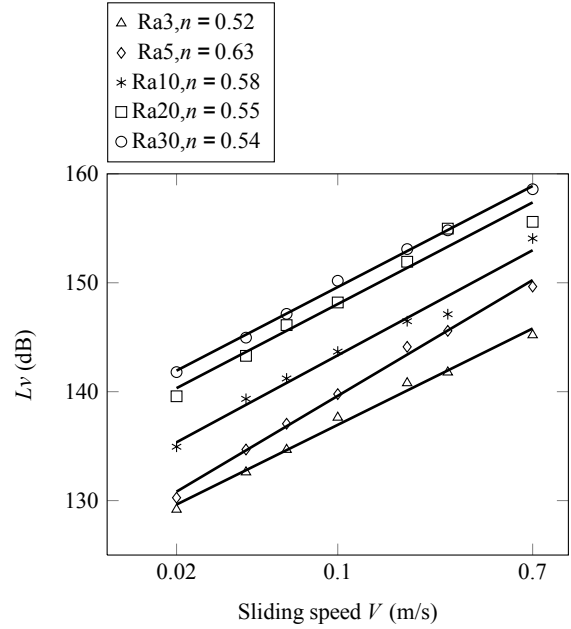}
\caption{Evolution of vibration level $Lv$ versus sliding speed $V$ for various surface roughness}
\label{Fi:Lv-V-2D}
\end{figure}
\begin{figure}[!ht]
\includegraphics{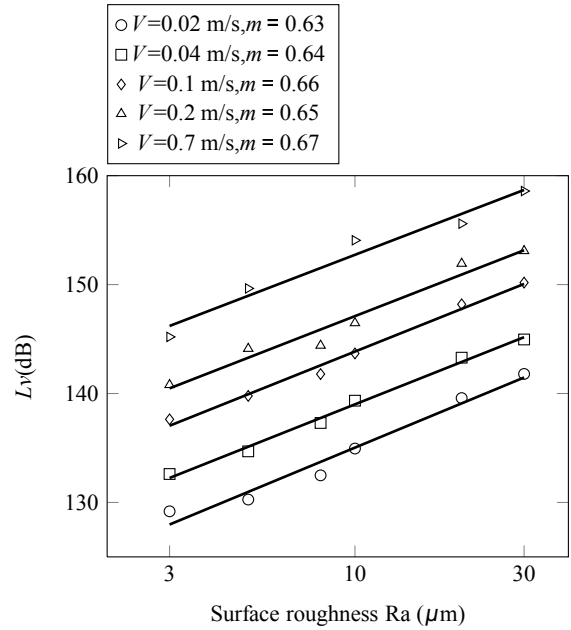}
\caption{Evolution of vibration level $Lv$ versus surface roughness $Ra$ for various velocity $V$}
\label{Fi:Lv-Ra-2D}
\end{figure}

The evolution of $Lv$ versus the sliding speed and the surface roughness are plotted in log-log scale in Figs.~\ref{Fi:Lv-V-2D} and~\ref{Fi:Lv-Ra-2D} respectively. The roughness noise depends simultaneously on $V$ and $Ra$ in agreement with experiments and with Eq.~(1). The exponents in this equation are determined from the figures. They are respectively  $0.63 \leqslant m \leqslant 0.67$ and $0.52 \leqslant n \leqslant 0.63$. These values are slightly smaller than the values found experimentally by Ben Abdelounis $m=0.8-1.16$ and $n=0.7-0.96$~\cite{Hou2010exp}. 

\begin{figure}[!ht]
\centering
\includegraphics{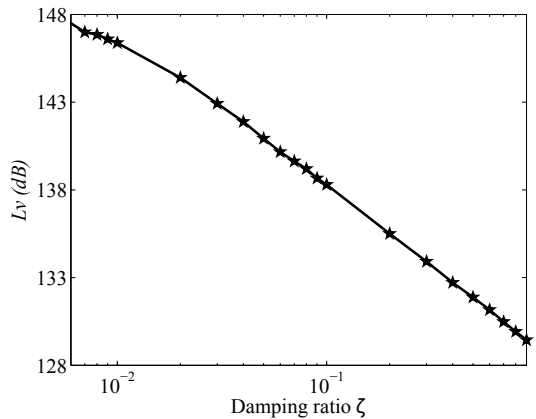}
\caption{Evolution of vibration level $Lv$ versus damping ratio $\zeta$. ($Ra$=10 $\mu$m, $V$= 0.1 m/s.)}
\label{Fi:damping-Lv}
\end{figure}

The viscous parameter is an important factor which affects strongly the vibration level. It is well-known in engineering that increasing dissipation (by adding a damping layer for instance) reduces sound level. Since dissipation is directly involved in the energy balance, a multiplication by 10 of the modal damping ratio results in a decrease by 10 dB of the vibrational level (see Fig.~\ref{Fi:damping-Lv}, on the example of $Ra$=10 $\mu$m and $V$=0.1 m/s, for our realistic problem).

Eventually, the CPU time is 9930 s (2.8 hours) per simulation. There are 6 rough profiles from Ra3 to Ra30 and 7 sliding speed from 0.02 m/s to 0.7 m/s which are used to calculate the evolution of the vibration level. Thus, the total CPU time for the realistic problem is around 116 hours.
\section{Conclusion}
\label{SE:Conclusion}
In this paper, we have focused on the vibration induced by the contact dynamics of rough surfaces. The motion of the system is governed by a partial differential equation in which two profiles are locally coupled by a contact force at some of the highest asperities. Solving this problem by an analytical approach would be difficult due to the randomness of surface and the non-linearity of the contact. 

A direct numerical simulation for the sliding contact between rough surfaces has been presented. The modal decomposition of the transverse vibration transforms the governing equation into a system of ordinary differential equation which is numerically solved by a central difference scheme. The contact is detected by the node-to-segment algorithm, and the contact forces are calculated by using either the penalty or Lagrange multipliers algorithm. This numerical approach was implemented in the software program RA2D. 

 Two validation examples show the accuracy and the rapidity of the program RA2D. The first example is a comparison with the analytical solution of a moving mass problem for  which a perfect agreement is observed. The second example is a comparison with the finite element method in a toy model consisting in a simple asperity problem. This validates the contact part of RA2D and highlights the influence of the main parameters such as the number of modes and the penalty coefficient.
  
The direct numerical simulation of a realistic problem have been presented at two length scales. First, the asperity-scale shocks can be determined from the time evolution of contact forces. It allows the characterisation of the probability distributions of the properties of shocks (duration, force, transferred energy). Second, the macroscale vibration level $Lv$ of the resonator can be obtained as the space and time average of the vibrational velocity of nodes in the simulation. The vibration level is found to be a linearly increasing function of the logarithm of both the surface roughness and sliding speed, in good agreement with experimental results from the literature. Our method can thus predict the evolution of the vibration level at macro scale by assessing the characteristics of shock between asperities at the micro scale. Furthermore, we can easily modify the input parameters in order to extend the results to other materials or dimension of solids. This numerical approach could be used for more complex systems such as wheel/rail or tyre/road.

A significant advantage of our method is its potential for drastic CPU time reduction. For instance, solving a realistic problem with a space step of 5 $\mu$m and a time step of 0.01 $\mu$s takes only a few hours of CPU time for a 2D simulation. This is to be compared with several days of CPU time using the finite element method. This is directly due to the modal truncation that we used. The number of kept modes is to be chosen according to the application considered. Here we have limited the simulation to the audio range because we focused on the friction noise. However, if small scale phenomena are investigated, higher frequency modes will be required in order to account for the local deformation of asperities. This will subsequently require a smaller time step to ensure stability of the numerical results, and thus a much longer CPU time. As a conclusion, we expect that, as far as friction noise is concerned, extending our method to 3D will make realistic 3D simulations possible with a reasonable (few days) CPU time.

  \renewcommand{\theequation}{A.\arabic{equation}}
  \setcounter{equation}{0}  

\FloatBarrier
\begin{acknowledgements}
This work was performed within the framework of the Labex CeLyA of Universite de Lyon, operated by the French National Research Agency (ANR-10-LABX-0060/ANR-11-IDEX-0007).\end{acknowledgements}
\bibliographystyle{spbasic}      

\end{document}